\documentclass{iau}
\usepackage{graphicx}
\newcommand{\lae}{\lower 2pt \hbox{$\, \buildrel {\scriptstyle <}\over {\scriptstyle
\sim}\,$}}

\begin{document}
\title{Radio Emission from the Bow Shock of G2}
\author[P Crumley and P Kumar]{P.~Crumley$^1$, P.~Kumar$^2$}

\affiliation{$^1$Physics Department, University of Texas at Austin,
  Austin, TX 78712\\ email: {\tt crumleyp@physics.utexas.edu}
  \\[\affilskip]$^2$Astronomy Department, University of Texas at
  Austin, Austin, TX 78712
}
\maketitle 
\pubyear{2013} 
\volume{303}   
\setcounter{page}{1} 
\jname{The GC: Feeding and Feedback in a
  Normal Galactic Nucleus} 
\editors{Editors} 

\abstract{The radio flux from the synchrotron emission of electrons
  accelerated in the forward bow shock of G2 is expected to have
  peaked when the forward shock passes close to the pericenter from
  the Galactic Center, around autumn of 2013. This radio flux is model
  dependent. We find that if G2 were to be a momentum-supported bow
  shock of a faint star with a strong wind, the radio synchrotron flux
  from the forward-shock heated ISM is well below the quiescent radio
  flux of Sgr A*. By contrast, if G2 is a diffuse cloud, the radio
  flux is predicted to be much larger than the quiescent radio flux
  and therefore should have already been detected or will be detected
  shortly. No such radiation has been observed to date. Radio
  measurements can reveal the nature of G2 well before G2 completes
  its periapsis passage.}

\section{Introduction}
G2, a spatially-extended red source, is on a nearly radial orbit
headed towards the super-massive black hole at the Galactic Center,
Sgr A* \cite[(Gillessen \etal\ 2012, Gillessen \etal\ 2013, Phifer
  \etal\ 2013)]{Gillessen12, Gillessen13, Phifer13}. As G2 plunges
towards Sgr A*, it is supersonic with a Mach number of around
2. Therefore, G2 drives a bow shock into the hot ISM which is expected to
accelerate electrons to relativistic energies. These high-energy
electrons then produce synchrotron radiation in the radio band
\cite[(Narayan \etal\ 2012, S{\c a}dowski \etal\ 2013b)]{Narayan12,
  Sadowski13b}. The synchrotron flux peaks when the forward shock
reaches periapsis, where the magnetic field is strongest. The forward
shock will reach perapsis $\sim$7 months before G2 center of mass
does, so the forward shock emission should have peaked around autumn of 2013
\cite[(S{\c a}dowski 2013a)]{Sadowski13a}. The magnitude of the radio
flux depends on how many electrons are swept up into the shock and
therefore on the cylindrical size and nature of G2.

The nature of G2 is undetermined. When first discovered,
\cite{Gillessen12} hypothesized that G2 was a pressure-confined,
non-self-gravitating gas cloud, due to the fact that the
Brackett-gamma (Br-$\gamma$) luminosity of G2 is not changing with
time, $L_{\rm Br-\gamma }\sim 2 \times 10 ^{- 3} L_\odot$ , and the
Br-$\gamma$ velocity dispersion is increasing in a manner that is well
fit by a gas cloud with a radius of $\sim 2 \times 10^{15}$ cm being
tidally sheared by Sgr A*. Alternatively, there is another class of
models where G2 contains a very faint stellar core that emits gas as
it falls towards Sgr A* \cite[(Murray-Clay \& Loeb 2012, Scoville \&
  Burkert 2013, Ballone \etal\ 2013)]{Murray12, Scoville13,
  Ballone13}. The ionized gas is then tidally sheared and is the
source of the Br-$\gamma$ radiation seen as G2. According to Scoville \& Burkert
(2013), the ionized gas that is the source of the Br-$\gamma$
radiation is located in the cold dense inner shock of a
momentum-supported bow shock between a stellar wind from a hidden,
TTauri star and the hot ISM.

\section{Methodology}
To calculate the number of electrons swept into the bow shock, we
assume ISM particle number density and temperature are inversely
proportional to the distance from Sgr A*. We calculate the size of the
bow shock if G2 is a shocked stellar wind by assuming it remains in
pressure equilibrium with the ISM. We use \cite{Ballone13} wind
properties, with a mass-loss rate of $\dot{M}_{\rm w} =
8.8\times10^{-8}\ M_{\odot}/ {\rm yr}$ and a velocity $v_{\rm w} =
50\ {\rm km/s}$. In the cloud model we assume the area is equal to
$\pi\times10^{30}\ {\rm cm^2}$, as suggested by \cite{Narayan12}. If
the cloud stays in pressure equilibrium, this area may be smaller
which would reduce the synchrotron flux in the cloud model
\cite[(Shcherbakov 2013)]{Roman13}. To calculate the synchrotron
flux, we extend the methodology of \cite{Sadowski13b} to the case
where the bow shock area is changing. Using \cite{Sadowski13b}
particle-in-cell simulations, we assume that 5\% of electrons are
accelerated into a power law with a starting Lorentz factor
\(\gamma=7.5kT/m_ec^2 + 1\), where $T$ is the unshocked ISM
temperature. We assume $P_{\rm mag} = \chi P_{\rm gas}$ , $\chi
\approx 0.1$

\section{Results}
\begin{figure}[ht!]
\begin{center}
\includegraphics[width=\textwidth]{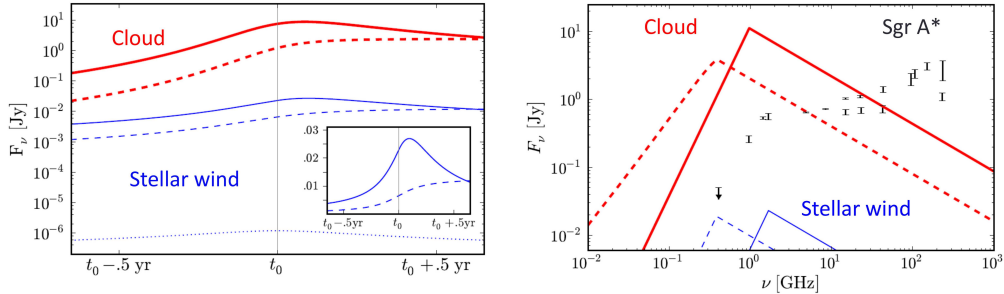}
\end{center}

\caption{{\bf Left:} The expected synchrotron flux at 1.4 GHz around
  pericenter passage of the forward shock, $t_0$, for different models
  of G2. Solid lines are the predicted fluxes when all of the
  accelerated electrons stay inside the bow shock. Dashed lines are
  the predicted fluxes when the electrons quickly leave the bow shock
  region after being accelerated. Thick red lines are the predicted
  fluxes assuming that G2 is a diffuse cloud and correspond to the
  larger prediction of the radio flux by \cite{Sadowski13b}. The lower
  blue lines are the predicted fluxes if G2 is a shocked stellar
  wind. The dotted blue line is the predicted flux from electrons
  accelerated in the inner, reverse shock. {\bf Left Inset:} A linear
  scale plot showing the flux at 1.4 GHz in Jy if G2 is a stellar
  wind.  {\bf Right:} Spectra at a time $t_0$ + 0.05 yr., where $t_0$
  is the periapsis crossing time of the forward shock for different
  models of G2. The color scheme is the same as left except the flux
  from the inner shock is omitted.
  The data points are radio fluxes measured during periods of
  inactivity of Sgr A* \cite[(Davies \etal\  1976, Falcke \etal\  1998, Zhao \etal\  2003)]{Davies76, Falcke98, Zhao03}.}
\label{fig1}
\end{figure}

We find that for both models of G2 considered, the radio flux of
forward shock of G2 peaks shortly after forward shock periapsis
crossing (see fig. \ref{fig1}). If G2 is a shocked stellar wind, the
radio flux will peak at a value around 0.02 Jy, which is far below the
radio flux of Sgr A*. This flux scales linearly with the wind
parameters $\dot{M}_{\rm w}$ and $v_{\rm w}$, as well as the fraction
of electrons that are accelerated.

\end{document}